\documentclass[utf8]{frontiersSCNS} 

\usepackage{url,hyperref,lineno,microtype,subcaption}
\usepackage[onehalfspacing]{setspace}

\def\keyFont{\fontsize{8}{11}\helveticabold }
\def\firstAuthorLast{S. Hekker} 
\def\Authors{S. Hekker\,$^{1,2,*}$}
\def\Address{$^{1}$Max Planck Institut f\"ur Sonnensystemforschung, G\"ottingen, Germany \\
$^{2}$Stellar Astrophysics Centre, Department of Physics and Astronomy, Aarhus University, Aarhus C, Denmark}
\def\corrAuthor{S. Hekker\\ Max Planck Institut f\"ur Sonnensystemforschung\\ Justus-von-Liebig-Weg 3\\ 37077 G\"ottingen, Germany}

\def\corrEmail{hekker@mps.mpg.de}

\begin{document}
\onecolumn
\firstpage{1}

\title[Scaling relation review]{Scaling relations for solar-like oscillations: a review} 

\author[\firstAuthorLast ]{\Authors} 
\address{} 
\correspondance{} 

\extraAuth{}

\maketitle

\begin{abstract}

The scaling relations for solar-like oscillations provide a translation of the features of the stochastic low-degree modes of oscillation in the Sun to predict the features of solar-like oscillations in other stars with convective outer layers. This prediction is based on their stellar mass, radius and effective temperature. Over time, the original scaling relations have been reversed in their use from predicting features of solar-like oscillations to deriving stellar parameters. Updates to the scaling relations as well as their reference values have been proposed to accommodate for the different requirements set by the change in their use. In this review the suggestions for improving the accuracy of the estimates of stellar parameters through the scaling relations for solar-like oscillations are presented together with a discussion of pros and cons of different approaches.

\tiny
 \keyFont{ \section{Keywords:} stellar pulsations, stellar parameters, solar-like oscillations, scaling relations, stellar mass, stellar radius} 
\end{abstract}

\section{Introduction}
With the advent of high resolution spectrographs (e.g. UCLES \citep{diego1990}, CORALIE \citep{queloz1999}, HARPS \citep{pepe2000}, UVES \citep{dekker2000} and SONG \citep{grundahl2007}) and dedicated space-based photometric missions (CoRoT \citep{michel1998}, \textit{Kepler} \citep{borucki2009}) the number of stars for which solar-like oscillations have been observed has increased by several orders of magnitude from the single case of the Sun \citep{leighton1962} to several hundreds to thousands \citep[e.g.,][]{hekker2009,chaplin2011,yu2018} over the last few decades.  Solar-like oscillations are stochastically excited by the turbulent convection in stars \citep[e.g.][]{goldreich1977,goldreich1988} with convective envelopes, i.e. in stars with effective temperatures below $\sim6700$~K. Effectively, some of the convective energy is transferred into energy of global oscillations, which reveal themselves as small amplitude oscillations at the stellar surface. As essentially all modes are excited the oscillation spectrum generally shows a clear pattern of overtones, with as a dominant feature the large frequency separation between modes of the same degree and consecutive radial order $\Delta\nu$. The oscillations are centred around a specific frequency (also called frequency of maximum oscillation power $\nu_{\rm max}$) with the (small) amplitudes of the oscillations decreasing away from this specific frequency.

In the 1980's and early 1990's, several groups attempted to observe solar-like oscillations in our brightest neighbouring stars such as Procyon, $\alpha$ Cen A, $\beta$ Hyi and $\epsilon$ Eri \citep{noyes1984,gelly1986,frandsen1987,brown1990,brown1991,innis1991,pottasch1992,bedford1993} to name a few. It was also at these times that the scaling relations (or asteroseismic scaling relations) for solar-like oscillations were first introduced. The main purpose of these relations was to predict the frequencies and amplitudes of the solar-like oscillations based on the known mass, radius, surface gravity and effective temperature of the target. This allowed for investigations as to whether the (null-)detections were genuine or due to limitations of the observations in terms of for instance signal-to-noise ratio and/or frequency resolution. 

An early suggestion for a scaling relation was presented by \citet{brown1991}. This scaling relation was based on the acoustic cut-off frequency ($\nu_{\rm ac}$), which is expected to scale as:
\begin{equation}
\nu_{\rm ac} \propto gT_{\rm eff}^{-\frac{1}{2}}
\label{nuac}
\end{equation}
with $g$ the surface gravity and $T_{\rm eff}$ the effective temperature. The predictions by \citet{brown1991} were based on the fact that the acoustic cut-off frequency is about 1.8 times the frequency at which the oscillation amplitudes in the Sun are largest. From this \citet{brown1991} predicted the location of the frequency of maximum oscillation power for Procyon to be around 1.0 mHz. 

\citet{kjeldsen1995} presented a dedicated study in which they predicted the amplitude (both velocity amplitude $v_{\rm osc}$ and luminosity amplitude $(\delta L/ L)_{\lambda}$ at wavelength $\lambda$), frequency of maximum oscillation power ($\nu_{\rm max}$) and large frequency separation ($\Delta\nu$) of other stars from scaling to the Sun, based on a linear adiabatic derivation. \citet{kjeldsen1995} formulated the scaling relations as follows:
\begin{equation}
v_{\rm osc} = \frac{L/L_{\odot}}{M/M_{\odot}}(23.4 \pm 1.4) \textrm{cm\,s}^{-1}
\label{ampv}
\end{equation}
\begin{equation}
(\delta L/L)_{\lambda} = \frac{L/L_{\odot} (4.7 \pm 0.3) \textrm{ ppm}}{(\lambda/550 \textrm{ nm})(T_{\rm eff}/5777 \textrm{ K})^2(M/M_{\odot})}
\label{amp}
\end{equation}
\begin{equation}
\Delta\nu_0 = (M/M_{\odot})^{\frac{1}{2}} (R/R_{\odot})^{-\frac{3}{2}} \Delta\nu_{\odot}
\label{dnu}
\end{equation}
\begin{equation}
\nu_{\rm max} = \frac{M/M_{\odot}}{(R/R_{\odot})^2 \sqrt{T_{\rm eff}/5777 \textrm{ K}}}\nu_{\rm max, \odot}
\label{numax}
\end{equation}
with $\Delta\nu_0$ the value of $\Delta\nu$ for radial (degree = 0) modes, $L$ luminosity, $M$ mass and $R$ radius. The $_{\odot}$ symbol indicates solar values, with  $\Delta\nu_{\odot} = 134.9$~$\mu \textrm{Hz}$ and $\nu_{\rm max, \odot} = 3.05\textrm{ mHz}$. Over the years, several authors have adopted different solar values based on internal calibrations from the analysis of a solar oscillation spectrum with the same method as applied to asteroseismic oscillation spectra. An overview of these values with references is provided in Table~\ref{solarvalues}.

The scaling relations provide decent estimates of the observed oscillations for a large range of stars. However, with the increase in the accuracy with which solar-like oscillations have been detected for a range of stars with different masses, metallicities and effective temperatures, the inherent shortcomings of such relations, i.e. they rely on a homologous stellar structure between the target star and the reference, have been apparent. Additionally, the use of the scaling relations has reversed from predicting oscillation features from known stellar parameters \citep[e.g.,][]{brown1991,kjeldsen1995} to estimating stellar parameters from the observed oscillations as per Eqs~\ref{Mrel} and \ref{Rrel} \citep[][were the first to apply this, to solar-type stars and red-giant stars, respectively]{stello2009R,kallinger2010}. This changed use of the scaling relations and our desire to obtain always more precise and accurate stellar parameters changed the accuracy and precision that we aim to reach with the scaling relations.
\begin{equation}
\frac{M}{\rm M_{\odot}} \simeq \left(\frac{\nu_{\rm max}}{\nu_{\rm max, \odot}}\right )^3 \left(\frac{\Delta\nu}{\Delta\nu_{\odot}}\right )^{-4}\left(\frac{T_{\rm eff}}{T_{\rm eff, \odot}}\right )^{3/2}
\label{Mrel}
\end{equation}
\begin{equation}
\frac{R}{\rm R_{\odot}} \simeq \left(\frac{\nu_{\rm max}}{\nu_{\rm max, \odot}}\right ) \left(\frac{\Delta\nu}{\Delta\nu_{\odot}}\right )^{-2}\left(\frac{T_{\rm eff}}{T_{\rm eff, \odot}}\right )^{1/2}
\label{Rrel}
\end{equation}

The amplitudes of the oscillations are related to the excitation and damping processes of the oscillations, which are still debated in the literature. Hence, the amplitude scaling relations (Eqs~\ref{ampv} and \ref{amp}) are not yet widely used to derive stellar parameters. On the other hand, the $\Delta\nu$ and $\nu_{\rm max}$ scaling relations (Eqs~\ref{dnu} and \ref{numax}) are now frequently used to determine stellar masses and radii (Eqs~\ref{Mrel} and \ref{Rrel}) and from these derive stellar ages. For this reason I focus here on the $\Delta\nu$ and $\nu_{\rm max}$ scaling relations.

\begin{table}
\begin{minipage}{\linewidth}
\caption{Overview of observed $\Delta\nu_{\odot}$ and $\nu_{\rm max, \odot}$ values as adopted in the literature.}
\label{solarvalues}
\centering
\begin{tabular}{ccl}
\hline
 $\Delta\nu_{\odot}$ [$\mu$Hz] & $\nu_{\rm max, \odot}$ [$\mu$Hz] & reference\\
\hline
134.9 & 3050 &  \citet{kjeldsen1995}\\
134.88 $\pm$ 0.04 & 3120 $\pm$ 5 & \citet{kallinger2010K}\\
134.9 & 3150 & \citet{chaplin2011}\\
135.1 $\pm$ 0.1 & 3090 $\pm$ 30 &  \citet{huber2011}\\
135.5 & 3050  & \citet{mosser2013}\\
134.9 $\pm$ 0.1 & 3060 $\pm$10  & \citet{hekker2013logg} (COR/EACF method)\\
135.03 $\pm$ 0.07 & 3140 $\pm$13  & \citet{hekker2013logg} (OCT method)\\
134.88 $\pm$ 0.04 & 3140 $\pm$ 5  & \citet{kallinger2014}\\
135.4 $\pm$ 0.3 & 3166 $\pm$ 6 & \citet{themessl2018}\\
\hline
\end{tabular}
\end{minipage}
\end{table}

\section{The $\Delta\nu$ and $\nu_{\rm max}$ scaling relations}
Here I first discuss the physical relation between stellar parameters and $\Delta\nu$ and $\nu_{\rm max}$, respectively. I subsequently present an overview of many of the validity tests and suggestions to adapt the scaling relations and/or the reference values which aim to improve the accuracy of the derived stellar parameters in chronological order. 

\subsection{Relation of $\Delta\nu$ and $\nu_{\rm max}$ with stellar parameters}
The $\Delta\nu$ scaling relation is physically justified as $\Delta\nu$ is in an asymptotic approximation equal to the inverse of the sound travel time through the star:
\begin{equation}
\Delta\nu = \left(2 \int_0^R \frac{dr}{c_s} \right)^{-1},
\label{dnuint}
\end{equation}
with $c_s$ the adiabatic sound speed. \citet{kjeldsen1995} showed that with estimates for internal values of the pressure and the temperature this results in $\Delta\nu \propto \sqrt{M/R^3}$, i.e. that the large frequency separation is directly proportional to the square root of the mean density of the star.

The $\nu_{\rm max}$ scaling relation has been defined empirically based on homology arguments with another typical dynamical timescale of the atmosphere, i.e. the acoustic cut-off frequency ($\nu_{\rm ac}$, see Eq.~\ref{nuac}). \citet{belkacem2011} aimed to provide a theoretical basis for the scaling between $\nu_{\rm max}$ and $\nu_{\rm ac}$. These authors indeed confirmed for stars other than the Sun that $\nu_{\rm max}$ corresponds to the plateau (depression) of the damping rates, as was already pointed out for the solar case by \citet{chaplin2008}. This combined with the suggestion by \citet{balmforth1992} that the plateau of the damping rate occurs when there is a resonance between the thermal time scale ($\tau$) and the modal frequency, \citet{belkacem2011} derived the resonance condition to be:
\begin{equation}
\nu_{\rm max} \simeq \frac{1}{2\pi \tau}.
\label{resonance}
\end{equation}
For a grid of models \citet{belkacem2011} found a close to linear relation between the thermal frequency $\tau^{-1}$ and $\nu_{\rm ac}$ with some dispersion related to the dispersion in mass. Hence, they concluded that the observed relation between $\nu_{\rm max}$ and $\nu_{\rm ac}$ is indeed the result of the resonance between $\nu_{\rm max}$ and $\tau^{-1}$, as well as the relation between $\tau^{-1}$ and $\nu_{\rm ac}$. \citet{belkacem2011} took this one step further to express this in thermodynamic quantities and found:
\begin{equation}
\nu_{\rm max} \propto \frac{1}{\tau} \propto \left( \frac{\Gamma_1^2}{\chi_{\rho}\Sigma} \right) \left( \frac{\mathcal{M}_a^3}{\alpha_{\rm MLT} }\right) \nu_{\rm ac},
\label{numaxscal}
\end{equation}
with $\mathcal{M}_a$ the Mach number, i.e. the ratio of the convective rms velocity $v_{\rm conv}$ to sound speed $c_s$, $\alpha_{\rm MLT}$ the mixing-length parameter, $\chi_{\rho} =(\partial \ln P/ \partial \ln \rho)_T$, $\Sigma = (\partial \ln \rho / \partial \ln T)_{\mu,P}$ and $\Gamma_1 = (\partial \ln P/ \partial \ln \rho)_{\rm ad}$ with $P$, $T$, $\rho$ and $\mu$ the pressure, temperature, density and mean molecular weight respectively. Finally, \citet{belkacem2011} stated that although the observed scaling between $\nu_{\rm max}$ and $\nu_{\rm ac}$ may not be obvious at first glance as $\nu_{\rm max}$ depends on the dynamical properties of the convective region while $\nu_{\rm ac}$ is a statistical property of the surface layers, the additional dependence on the Mach number resolves this paradox.

Together the $\Delta\nu$ and $\nu_{\rm max}$ scaling relations (Eqs~\ref{dnu} and \ref{numax}) can be rewritten to provide stellar masses and radii (Eqs~\ref{Mrel} and \ref{Rrel}). This path way of deriving stellar masses and radii is now widely in use. Hence, the $\Delta\nu$ and $\nu_{\rm max}$ scaling relations are discussed here together.

\subsection{Validity tests \& suggested improvements}
After some initial general investigations in the validity of the $\Delta\nu$ scaling relation by \citet{stello2009}, \citet{bruntt2010} and \citet{basu2010}, \citet{white2011} were the first to carry out an in depth study on how accurately the relation in Eq.~\ref{white} is followed by models:
\begin{equation}
\rho \approx \left(\frac{\Delta\nu}{\Delta\nu_{\odot}}\right)^2 \rho_{\odot},
\label{white}
\end{equation}
with $\rho$ and $\rho_{\odot}$ the mean density of the star and the Sun, respectively. In their work, \citet{white2011} computed $\Delta\nu$ from a linear (Gaussian-weighted) least squares fit to the frequencies of radial modes. Throughout the paper, I will refer to $\Delta\nu$ derived in a similar way as $\Delta\nu_{\rm freq}$. Using the same approach \citet{white2011} computed $\Delta\nu_{\odot}$~=~135.99~$\mu$Hz derived from a fit to frequencies of the standard solar model, model S of \citet{JCD1996}.

\citet{white2011} showed that deviations from the scaling relation exist in models and that these are predominantly a function of effective temperature. For stars with temperatures in the range 4700~K to 6700~K and masses larger than $\sim$~1.2~M$_{\odot}$, these authors suggested a variation of the scaling relation of the form:
\begin{equation}
\frac{\rho}{\rho_{\odot}} =\left(\frac{\Delta\nu}{\Delta\nu_{\odot}}\right)^2 (f(T_{\rm eff}))^{-2},
\label{white1}
\end{equation}
where
\begin{equation}
f(T_{\rm eff}) = -4.29\left(\frac{T_{\rm eff}}{10^4\textrm{ K}}\right)^2+4.84\left(\frac{T_{\rm eff}}{10^4\textrm{ K}}\right)-0.35.
\label{white2}
\end{equation}
According to \citet{white2011}, metallicity has little effect except for red giants, for which there is a slight dependence. Furthermore, they noted that their function (Eqs~\ref{white1} and \ref{white2}) is based on models and the so-called surface effect (a frequency-dependent offset between observed and modelled frequencies that affects $\Delta\nu$) is not accounted for. Nevertheless, they recommended Eq.~\ref{white1} (or Eq.~\ref{white}) to be used with the observed value of $\Delta\nu_{\odot} = 135.0$~$\mu$Hz.

Subsequently, \citet{huber2012} compared the radii of stars measured from asteroseismic scaling relations with radii measured from interferometry. They obtained excellent agreement within the observational uncertainties. They furthermore showed that asteroseismic radii of main-sequence stars are accurate to $\leq$4 per cent. At about the same time \citet{silvaaguirre2012} used the oscillation data and multi-band photometry to derive stellar parameters in a self-consistent manner coupling asteroseismic analysis with the Infra Red Flux Method (IRFM). They showed an overall agreement of 4 per cent with \textit{Hipparcos} parallaxes, a mean difference in $T_{\rm eff}$ of less than 1 per cent and agreement within 5 per cent for the angular diameters. Despite these encouraging results, \citet{silvaaguirre2012} warned for systematics either due to reddening or metallicity, or due to observational uncertainties.

Following \citet{stello2009} and \citet{kallinger2010}, there have been many attempts to use the scaling relations to determine stellar masses and radii, either directly or from grid-based modelling \citep[e.g.][]{gai2011}. In one of these works  \citet{miglio2012}  explicitly addressed the fact that stars on the red-giant branch (RGB) have an internal temperature (hence sound speed) distribution different from that of stars in the core helium burning phase (CHeB). They found that an CHeB model has a mean $\Delta\nu$ that is about 3.3 per cent larger than an RGB model, despite having the same mean density. This difference is due to the fact that the sound speed in the CHeB model is on average higher (at a given fractional radius) than that of the RGB model, mostly due to the different temperature profiles. This effect is largest in the region below the boundary of the helium core in the RGB model, though the near-surface regions ($r/R \ge 0.9$) also contribute about 0.8 per cent. 
Based on this finding \citet{miglio2012} suggested that a relative correction has to be considered when dealing with CHeB stars and RGB stars. This relative correction is expected to be mass-dependent and to be larger for low-mass stars, which have significantly different internal structures when ascending the RGB compared to when they are in the CHeB.

\citet[][see also \citet{mosser2013a,mosser2013b}]{mosser2013} made an explicit link between the asymptotic spacing ($\Delta\nu_{\rm as}$, the value of $\Delta\nu$ as defined in Eq.~\ref{dnuint}) and the observed spacing ($\Delta\nu_{\rm obs}$), where $\Delta\nu_{\rm obs}$ is defined as the difference in observed frequencies of radial modes. \citet{mosser2013} linked $\Delta\nu_{\rm as}$ with $\Delta\nu_{\rm obs}$ in the following way:
\begin{equation}
\Delta\nu_{\rm as} = \Delta\nu_{\rm obs} \left ( 1+\frac{n_{\rm max} \alpha_{\rm obs}}{2} \right),
\label{dnuas}
\end{equation}
with $\alpha_{\rm obs}$ the curvature and $n_{\rm max}$ a dimensionless value of $\nu_{\rm max}$ defined as $n_{\rm max} = \nu_{\rm max} / \Delta\nu_{\rm obs}$. By taking into account the curvature, it is possible to correct the observed value of $\Delta\nu$ and derive its asymptotic counter part, which leads to more accurate asteroseismic estimates of the stellar mass and radius \citep[see also][]{belkacem2013dnu}. \citet{mosser2013} stated that in case the asymptotic values are used (together with the solar values as listed in Table~\ref{solarvalues}) no correction has to be applied. If the observed values are used, then corrections up to 7.5 per cent and 2.5 per cent in mass and radius should be applied. Alternatively, \citet{mosser2013} suggested to use in combination with the observed $\Delta\nu$ and $\nu_{\rm max}$ more general reference values, i.e. $\Delta\nu_{\rm ref}$, instead of the solar reference values. So for stars other than the Sun, they suggested these new calibrated references to be $\Delta\nu_{\rm ref} =138.8$~$\mu$Hz and $\nu_{\textrm{max, ref}} = 3104$~$\mu$Hz.

In response to the work by \citet{mosser2013}, \citet{hekker2013} investigated whether the differences between observable oscillation parameters and their asymptotic estimates are indeed significant. Based on stellar models they found that the extent to which the atmosphere is included in the model is a key parameter. Considering a larger extension of the atmosphere beyond the photosphere reduces the difference between the asymptotic and observable values of the large frequency separation. Hence, \citet{hekker2013} cautioned that the corrections proposed by \citet{mosser2013} may be overestimated.

\citet{epstein2014} tested masses obtained from asteroseismic scaling relations against masses of metal-poor ([Fe/H]~$<-1$) stars. Based on the fact that the nine stars (6 halo stars and 3 thick disc stars) in their study can not be younger than 8~Gyr combined with models with a normal (near-primordial) helium abundance provided a range of theoretically allowed masses of between roughly 0.8 and 0.9~M$_{\odot}$. The masses obtained by (uncorrected) scaling relations are overestimated by about 16 per cent. This overestimate reduced by including corrections to the reference values of the scaling relations from \citet{kallinger2010,white2011,mosser2013}, though they did not mitigate the problem fully. This prompted \citet{epstein2014} to call for further investigations into the metallicity dependence of the $\Delta\nu$ scaling relation and the impact of the $\nu_{\rm max}$ scaling relation on mass estimates.

\citet{coelho2015} performed tests on how well the oscillations of cool main-sequence  and subgiant stars adhere to the relation between $\nu_{\rm max}$ and the cut-off frequency for acoustic waves in an isothermal atmosphere. The results by \citet{coelho2015} based on a grid-based modelling approach ruled out departures from the classic $\nu_{\rm max}$ scaling relation at the level of $\sim$1.5 per cent over the full range in $T_{\rm eff}$ (5600~K~$< T_{\rm eff} < 6900$~K) that they tested for. \citet{coelho2015} stated that there is some uncertainty concerning the absolute calibration of the scaling relation, though any variation with $T_{\rm eff}$ is small, resulting in a limit similar to the $\sim$1.5 per cent level.

\citet{brogaard2016} concluded in their ongoing investigations of the asteroseismic scaling relations in open cluster stars and binaries that they are accurate to within their uncertainties for giant stars. They stated that this is the case as long as corrections to the reference values of the $\Delta\nu$ scaling relation are calculated and applied along the lines of \citet{miglio2013} whom considered a 5 per cent systematic uncertainty on the radius determination to account for inaccuracies in the scaling relations. \citet{brogaard2016} noted that asteroseismic $\log g$ values are extremely consistent with their independent measurements which implies that the scaling for $\nu_{\rm max}$ is reliable.

\citet{sharma2016} proposed a correction factor $f_{\Delta\nu}$ defined as:
\begin{equation}
f_{\Delta\nu} = \left(\frac{\Delta\nu}{\Delta\nu_{\odot}} \right) \left( \frac{\rho}{\rho_{\odot}} \right) ^{-0.5},
\label{correctsharma}
\end{equation}
with $\Delta\nu_{\odot}=135.1$~$\mu$Hz. The value of $f_{\Delta\nu}$ was determined for a grid of models with $-3.0$~dex~$<$~[Fe/H]~$<$~0.4~dex and 0.8~M$_{\odot}$~$<$~$M$~$<$~4.0~M$_{\odot}$ following the same approach as \citet{white2011} to derive $\Delta\nu_{\rm freq}$ for each model in a way to mimic the way $\Delta\nu$ is measured from data. The value of $f_{\Delta\nu}$ was obtained by \citet{sharma2016} along each stellar track ranging from the zero-age main sequence until the end of helium-core burning. These results were combined in a grid, for which they computed the correction factor for each synthetic star through an interpolation and they corrected $\Delta\nu$ based on this factor. Additionally, \citet{sharma2016} also applied a correction to the $\nu_{\rm max}$ scaling relation of $f_{\nu_{\rm max}}=1.02$ to improve the agreement between the models and observations.

\citet{guggenberger2016} tackled the issue of the dependence of the $\Delta\nu$ reference on both effective temperature and [Fe/H] by fitting a  $T_{\rm eff}$ - [Fe/H] dependent reference function through a set of models spanning $-1.0$~dex~$<$~[Fe/H]~$<$~0.5~dex and 0.8~M$_{\odot}$~$<$~$M$~$<$~2.0~M$_{\odot}$. Based on the variations in the ratio of the value of $\Delta\nu$ from scaling relations with solar values to values of $\Delta\nu_{\rm freq}$ obtained from the differences between radial oscillation modes as a function of $T_{\rm eff}$ in stellar models, this reference function has the following shape:
\begin{equation}
\Delta\nu_{\rm ref} = Ae^{\lambda T_{\rm eff} / 10^4 \textrm{K}}(cos(\omega T_{\rm eff} / 10^4\textrm{K} + \phi))+B,
\label{refgug16}
\end{equation}
with
\begin{eqnarray}
A=0.64\textrm{[Fe/H]} + 1.78\textrm{ } \mu \textrm{Hz},\\
\lambda=-0.55\textrm{[Fe/H]} + 1.23,\\
\omega=22.12 \textrm{ rad\,K}^{-1},\\
\phi=0.48\textrm{[Fe/H]} + 0.12,\\
B=0.66\textrm{[Fe/H]} + 134.92\textrm{ }\mu \textrm{Hz},
\end{eqnarray}
and was calibrated for stars in different evolutionary states including (end of) main-sequence stars, subgiants and cool red giants down to $\nu_{\rm max} = 6$~$\mu$Hz.
Similar to \citet{white2011} this reference function was developed on models and does not include the surface correction. Nevertheless, \citet{guggenberger2016} showed that this reference function allows masses and radii to be recovered through asteroseismic scaling relations with an accuracy of 5 per cent and 2 per cent, respectively. For this they used $\nu_{\textrm{max, ref}} = \nu_{\rm max,\odot} = 3050$~$\mu$Hz.

\citet{gaulme2016} subsequently tested for 10 red-giant stars the masses and radii obtained from the asteroseismic scaling relations against masses and radii obtained from the orbital solutions of spectroscopic eclipsing binaries. These authors found that the asteroseismic scaling relations overestimate the radii by about 5 per cent on average and the masses by about 15 per cent on average, while using the $\Delta\nu$ scaling relation where the curvature was included as proposed by \citet{mosser2013}. \citet{gaulme2016} also tested both the original scaling relations \citep{kjeldsen1995} as well as other reference values \citep{kallinger2010K,chaplin2011,guggenberger2016} and corrections to the scaling relations \citep{sharma2016}, with similar or worse results. \citet{gaulme2016} noted that another culprit in the scaling relations is the effective temperature, i.e., overestimated temperatures can lead to overestimated values for the scaling law masses and radii. Indeed, when \citet{gaulme2016} decreased their effective temperatures by 100~K the asteroseismic masses and radii decreased by 3.1 per cent and 1.0 per cent, respectively.

\citet{yildiz2016} investigated the impact of the assumption that the first adiabatic exponent ($\Gamma_1$) and mean molecular weight ($\mu$) are assumed to be constant at the stellar surface for the purpose of deriving the scaling relations. \citet{yildiz2016} found that depending on the effective temperature,  $\Gamma_1$ changes significantly in the near surface layers of solar-like stars. Henceforth, they found that the ratio of the mean large frequency separation to square root of mean density is a linear function of $\Gamma_1$. Additionally, they also included the $\Gamma_1$ dependence into the $\nu_{\rm max}$ scaling relation. The relations to determine stellar mass and radius as proposed by \citet{yildiz2016} are as follows:
\begin{equation}
\frac{M}{M_{\odot}} = \frac{(\nu_{\rm max} / \nu_{\rm max \odot})^3}{(\Delta\nu/\Delta\nu_{\odot})^4} \left( \frac{T_{\rm eff}}{T_{\rm eff \odot}} \frac{\Gamma_{1 \odot}}{\Gamma_1} \right)^{\frac{3}{2}} \frac{f_{\Delta\nu}^4}{f_{\nu}^3},
\label{massyildiz}
\end{equation}
\begin{equation}
\frac{R}{R_{\odot}} = \frac{(\nu_{\rm max} / \nu_{\rm max \odot})}{(\Delta\nu/\Delta\nu_{\odot})^2} \left( \frac{T_{\rm eff}}{T_{\rm eff \odot}} \frac{\Gamma_{1 \odot}}{\Gamma_1} \right)^{\frac{1}{2}} \frac{f_{\Delta\nu}^2}{f_{\nu}},
\label{massyildiz}
\end{equation}
with
\begin{equation}
f_{\Delta\nu} = 0.430 \frac{\Gamma_1}{\Gamma_{1\odot}} + 0.570,
\end{equation}
\begin{equation}
f_{\nu} = 0.470 \frac{\Gamma_{1\odot}}{\Gamma_1} + 0.530.
\end{equation}

Following \citet{yildiz2016}, \citet{viani2017} examined the $\nu_{\rm max}$ scaling relation taking into account that the first adiabatic exponent ($\Gamma_1$) and mean molecular weight ($\mu$) are not constant at the stellar surface. Based on models they found that the largest source of the deviation in the $\nu_{\rm max}$ scaling relation is the neglect of the mean molecular weight ($\mu$) and $\Gamma_1$ terms when approximating the acoustic cut-off frequency. \citet{viani2017} proposed the following relation to be used:
\begin{equation}
\frac{\nu_{\rm max}}{\nu_{\rm max,\odot}} = \left( \frac{M}{M_{\odot}} \right) \left( \frac{R}{R_{\odot}} \right)^{-2} \left( \frac{T_{\rm eff}}{T_{\rm eff,\odot}} \right)^{-\frac{1}{2}} \left( \frac{\mu}{\mu_{\odot}} \right)^{\frac{1}{2}} \left( \frac{\Gamma_1}{\Gamma_{1,\odot}} \right)^{\frac{1}{2}}.
\label{numaxviani}
\end{equation}
\citet{viani2017} noted that the deviations in the scaling relations cause systematic errors in estimates of $\log g$, mass and radius. The errors in $\log g$ are however well within errors caused by data uncertainties and are therefore not a big cause for concern, except at extreme metallicities.

Following on from the $T_{\rm eff}$ - [Fe/H] dependent reference function, \citet{guggenberger2017} performed symbolic regression, i.e. they let both the functional form as well as the parameters vary to obtain a best fit, to mitigate the mass dependence of $\Delta\nu_{\rm ref}$ for stars with 5~$\mu$Hz~$<$~$\nu_{\rm max}$~$<$~170~$\mu$Hz. Essentially, two functions were presented: one based directly on the $\Delta\nu$ derived from the models in a way to mimic the observations and one after applying the reference function of \citet{guggenberger2016} (see Eq.~\ref{refgug16}). These functions take the following from:
\begin{equation}
\Delta\nu_{\rm ref} = A_1+A_2 \times M+\frac{A_3}{\nu_{\rm max}}+A_4 \times \sqrt{\nu_{\rm max}}-A_5 \times \nu_{\rm max}-A_6 \times \textrm{[Fe/H]},
\label{refgug17_1}
\end{equation}
and for the residuals of Eq.~\ref{refgug16}:
\begin{equation}
\Delta\nu_{\rm ref, residuals} = B_1 \times M+B_2 \times \nu_{\rm max}+\frac{B_3 \times M -B_4 \times \textrm{[Fe/H]}}{\nu_{\rm max}}-B_5 - B_6 \times M \times \nu_{\rm max},
\label{refgug17_2}
\end{equation}
where the values of the parameters and units are listed in Table~\ref{tab:param}. As the mass $M$ is included in these functions, they have to be applied in an iterative manner. In the range 5~$\mu$Hz~$<$~$\nu_{\rm max}$~$<$~170~$\mu$Hz the reference functions Eqs~\ref{refgug17_1} and \ref{refgug17_2} improve mass and radius determinations by 10 per cent and 5 per cent respectively (compared to using a solar reference). This is true in the limit of ideal data obtained from canonical stellar models and without including a surface effect. \citet{guggenberger2017} noted that Eqs~\ref{refgug17_1}, \ref{refgug17_2} as well as \ref{refgug16} do not have a physical meaning. However, they do represent an empirical fit optimised to the data obtained from stellar models that include canonical stellar physics. 

\begin{table}
	\centering
	\caption{Parameters with their units of the functions in Eqs~\ref{refgug17_1} and \ref{refgug17_2}.}
	\label{tab:param}
	\begin{tabular}{lrrclrr} 
		\hline
$A_1$	&	124.72	&  $\mu$Hz &	\  \	&	$B_1$	&	1.88	&  $\mu$Hz/M$_\odot$	\\
$A_2$	&	2.23	&  $\mu$Hz/M$_\odot$ &	\  \	&	$B_2$	&	0.02  &  -	\\	
$A_3$	&	17.61	&  $\mu$Hz$^2$ & \  \ &	$B_3$	&	5.14 & $\mu$Hz$^2/$M$_\odot$		\\
$A_4$	&	0.73	&  $\sqrt{\mu \rm Hz}$  & \  \ &	$B_4$	&	10.90 &  $\mu$Hz$^2$	\\
$A_5$	&	0.02	&  - &	\  \	&	$B_5$	&	3.69 &  $\mu$Hz		\\
$A_6$	&	0.93	&  $\mu$Hz &	\  \	&	$B_6$	&	0.01 & M$_{\odot}^{-1}$	\\	
		\hline
	\end{tabular}
\end{table}

\citet{serenelli2017} formulated a calibration factor to account for the surface effects in cases where $\Delta\nu$ in stellar models is computed from theoretical frequencies \citep[e.g.,][]{sharma2016,rodrigues2017}. The advantage of relying on $\Delta\nu_{\rm freq}$ computed from theoretical frequencies is that it captures deviations from the pure scaling relation due to the detailed structure of stellar models \citep[e.g.,][]{belkacem2013dnu}. However, the underlying theoretical frequencies are affected by poor modelling of stellar atmospheres and the neglect of non-adiabatic effects in the outer most layers \citep{rosenthal1999}, i.e. the surface effect. Therefore, the $\Delta\nu_{\rm freq}$ from solar models is about 1 per cent larger than the observed $\Delta\nu_{\odot}$. This difference implies that stellar model grids that rely on $\Delta\nu_{\rm freq}$ computed from theoretical frequencies will not be able to reproduce a solar model unless it is rescaled to match $\Delta\nu_{\odot}$. The calibration factor $f_{\rm cal}$ to rescale $\Delta\nu_{\rm freq}$ to $\Delta\nu_{\odot}$ suggested by \citet{serenelli2017} is as follows:
\begin{equation}
f_{\rm cal} = \frac{\Delta\nu_{\odot}}{\Delta\nu_{\textrm{freq,SM}}},
\label{coraldo}
\end{equation}
where SM means solar model. Such a rescale has been applied by \citet{serenelli2017} to the full grid of stellar models used to compute stellar parameters.

In a similar approach as \citet{gaulme2016}, \citet{brogaard2018} and \citet{themessl2018} tested the asteroseismic masses and radii against masses and radii obtained from binary orbits for three eclipsing binary systems each (one system in overlap). Both studies found that asteroseismic scaling relations without corrections to the $\Delta\nu$ scaling relations would overestimate the masses and radii. However, by including the theoretical correction factors ($f_{\Delta\nu}$) according to \citet{rodrigues2017}\footnote{\citet{rodrigues2017} implemented a similar interpolation scheme in their models as \citet{sharma2016}. They also experimented with the impact of the period spacing $\Delta P$ on the mass and radius determination, though that is beyond the scope of this review.}, \citet{brogaard2018} reached general agreement between dynamical and asteroseismic mass estimates, and no indications of systemic differences at the level of precision of the asteroseismic measurements. In the same vein, \citet{themessl2018} proposed an empirical reference value for $\Delta\nu$ ($\Delta\nu_{\textrm{ref,emp}}$) that is consistent with the corrections by \citet{guggenberger2016} while also including surface effects as computed for the same set of stars by \citet{ball2018}. \citet{themessl2018} presented the following value:
\begin{equation}
\Delta\nu_{\textrm{ref,emp}} = 130.8 \pm 0.9\textrm{ } \mu \textrm{Hz},
\label{refthemessl}
\end{equation}
with a consistent solar reference for $\nu_{\rm max}$ of 3137~$\pm$~45~$\mu$Hz.
Both the studies by \citet{brogaard2018} and \citet{themessl2018} indicated that this is just a start and that there is a need for a large high-precision sample of eclipsing spectroscopic binaries (eSB2) covering a range in mass, metallicity and stellar evolution to further test the masses and radii of solar-like oscillators determined through scaling relations.

\citet{kallinger2018} devised non-linear seismic scaling relations based on six known eSB2 systems selected from \citet{gaulme2016,themessl2018,brogaard2018}. By comparing $\nu_{\rm max}$ to $g_{\rm dyn} / \sqrt{T_{\rm eff}}$, where $g_{\rm dyn}$ is the surface gravity derived from the dynamical solution of the red-giant components in the eSB2 systems, they obtained a reference value for $\nu_{\rm max}$ for RGB stars with 20~$\mu$Hz~$<$~$\nu_{\rm max}$~$<$~80~$\mu$Hz of $\nu_{\textrm{max,ref,RGB}}$~=~3245~$\pm$~50~$\mu$Hz. For a more general approach \citet{kallinger2018} fitted
\begin{equation}
\frac{g_{\rm dyn}}{\sqrt{T_{\rm eff}}} = \left( \frac{\nu_{\rm max}}{\nu_{\rm max,\odot}} \right)^{\kappa},
\label{refnumaxkal}
\end{equation}
in which $\nu_{\rm max,\odot}$~=~3140~$\pm$~5~$\mu$Hz \citep{kallinger2014}. \citet{kallinger2018} found $\kappa = 1.0080\pm0.0024$.
For the large frequency separation, \citet{kallinger2018} found a similar situation. The average of the six stars provides a reference value $\Delta\nu_{\textrm{ref,RGB}}$ of $133.1\pm1.3$~$\mu$Hz. However they found more statistical evidence for the function:
\begin{equation}
\Delta\nu=\Delta\nu_{\rm ref} \cdot \sqrt{\overline{\rho}_{\rm dyn}} = \Delta\nu_{\odot}[1-\gamma\log^2(\Delta\nu/\Delta\nu_{\odot})]\cdot\sqrt{\overline{\rho}_{\rm dyn}},
\label{refdnukal}
\end{equation}
with $\gamma=0.0043\pm0.0025$ when using the average frequency spacing of the three central radial orders (local $\Delta\nu$ or $\Delta\nu_{\rm c}$) and a local solar value $\Delta\nu_{\rm c,\odot} = 134.89\pm0.04$~$\mu$Hz, or $\gamma=0.0085\pm0.0025$ when including a curvature and glitch correction (indicated with $\Delta\nu_{\rm cor}$) and a corrected solar value $\Delta\nu_{\rm cor,\odot} = 135.08\pm0.04$~$\mu$Hz. \citet{kallinger2018} noted that the latter solution should be preferred over the local or average value of $\Delta\nu$.

\citet{ong2019} derived an asymptotic estimator for the large frequency separation that captures most of the variations in the scaling relation with a single expression and thereby return estimates of $\Delta\nu$ that are considerably closer to the observed value than the traditional estimator, without any ambiguity as to the outer turning point of the relevant integral \citep[see][]{hekker2013}. They derived a new expression for $\Delta\nu$ by using a more accurate description of the WKB\footnote{One of the most useful techniques for studying wave-like solutions of ordinary linear differential equations of second order: namely the so-called Liouville-Green expansion combined with the method of Jeffreys for connecting solutions across turning points. See \citet{gough2007} for more details.} expression of the first-order asymptotic theory of p modes in which a more detailed asymptotic analysis (i.e., not setting terms to zero prematurely before performing the WKB analysis) was used \citep{deubner1984}. Following a Taylor expansion \citet{ong2019} derived:
\begin{equation}
\Delta\nu \sim \left(2 \int^{r_2}_{r_1} \frac{dr}{c_s} \frac{1}{\sqrt{1-\frac{\omega_{\rm ac}^2}{\omega^2}}} \right)^{-1},
\label{dnuong}
\end{equation}
in which $\omega = 2\pi \nu$ is the angular frequency and $\omega_{\rm ac}$ the angular acoustic cut-off frequency:
\begin{equation}
\omega^2_{\rm ac} = \frac{c_s^2}{4H^2} \left(1-2\frac{dH}{dr} \right),
\end{equation}
with $H$ the density scale height.
\citet{ong2019} showed that in this prescription the turning points of the integral emerge naturally from the theoretical formulation and do not suffer any ambiguity independent of the choice of model atmosphere or modifications to the model metallicity. The only precaution is that the integral expression (Eq.~\ref{dnuong}) becomes singular at some point during the main-sequence turn-off, which is ultimately a consequence of the failure of the WKB regime. \citet{ong2019} showed that these singular points occur during a transition between two extreme regimes of asymptotic behaviour providing theoretical justification for separately calibrated scaling relations for stars at different evolutionary stages.

Finally, \citet{bellinger2019b} used the \textit{Kepler} Ages \citep{silvaaguirre2015,davies2016} and LEGACY samples \citep{lund2017,silvaaguirre2017} to investigate the scaling relations for main-sequence stars. \citet{bellinger2019b} used the masses and radii from the Stellar Parameters in an Instant (SPI) method \citep{bellinger2016} as provided by \citet{bellinger2019a} to provide the following functions:
\begin{equation}
\frac{M}{M_{\odot}} = \left( \frac{\nu_{\rm max}}{\nu_{\rm max,\odot}} \right)^{0.975} \left( \frac{\Delta\nu}{\Delta\nu_{\odot}} \right)^{-1.435} \left( \frac{T_{\rm eff}}{T_{\rm eff,\odot}} \right)^{1.216} \exp\left(\textrm{[Fe/H]}\right)^{0.270},\\
\label{functionbel_M}
\end{equation}
\begin{equation}
\frac{R}{R_{\odot}} = \left( \frac{\nu_{\rm max}}{\nu_{\rm max,\odot}} \right)^{0.305} \left( \frac{\Delta\nu}{\Delta\nu_{\odot}} \right)^{-1.129} \left( \frac{T_{\rm eff}}{T_{\rm eff,\odot}} \right)^{0.312} \exp\left(\textrm{[Fe/H]}\right)^{0.100},\\
\label{functionbel_R}
\end{equation}
\begin{equation}
\frac{\tau}{\tau_{\odot}} = \left( \frac{\nu_{\rm max}}{\nu_{\rm max,\odot}} \right)^{-6.556} \left( \frac{\Delta\nu}{\Delta\nu_{\odot}} \right)^{9.059} \left( \frac{\delta\nu}{\delta\nu_{\odot}} \right)^{-1.292} \left( \frac{T_{\rm eff}}{T_{\rm eff,\odot}} \right)^{-4.245} \exp\left(\textrm{[Fe/H]}\right)^{-0.426},
\label{functionbel_tau}
\end{equation}
with $\nu_{\rm max,\odot} = 3090\pm 30$~$\mu$Hz, $\Delta\nu_{\odot}=135.1 \pm 0.1$~$\mu$Hz \citep{huber2011}, $T_{\rm eff,\odot} = 5772.0 \pm 0.8$~K \citep{prsa2016}; $\delta\nu$ is the small frequency separation between modes of degree 0 and 2 with $\delta\nu_{\odot} = 8.957 \pm 0.059$~$\mu$Hz \citep[based on data from][]{davies2014} and $\tau$ is age with $\tau_{\odot} = 4.569 \pm 0.006$~Gyr \citep{bonanno2015}.
\citet{bellinger2019b} stated that Eqs~\ref{functionbel_M}, \ref{functionbel_R} and \ref{functionbel_tau} yield uncertainties of 0.032~M$_{\odot}$ (3.3 per cent), 0.011~R$_{\odot}$ (1.1 per cent) and 0.56~Gyr (12 per cent) for mass, radius and age, respectively.


\section{Discussion}
The suggestions to improve the accuracy of the stellar parameters derived from the $\Delta\nu$ and $\nu_{\rm max}$ scaling relations as presented above focus on different aspects and follow different approaches, which all have pros and cons. The determination of alternative reference values \citep{mosser2013,themessl2018} or reference functions \citep{white2011,guggenberger2016,guggenberger2017} have the advantage of direct applicability to observed data without any use of models. The drawback is that the values or functions may not capture all dispersions in, for instance, mass, metallicity or temperature. Furthermore, the reference values and functions are derived for a certain parameter space or on stars in a certain parameter space, and hence, they will be most reliable in that parameter space.

When using models, a correction factor implemented throughout a grid \citep{sharma2016,rodrigues2017,serenelli2017} or the inclusion of $\Gamma_1$ and $\mu$ \citep{yildiz2016,viani2017} will allow to mitigate such dispersions. However, one has to rely on stellar models, and the physics included in the models. Additionally, the surface effect has to be accounted for in any comparison between models and observed data \citep{serenelli2017}. 

The approach of altering the shape of the scaling relations by including alternative exponents or non-linear terms \citep{kallinger2018,bellinger2019b} provides accurate stellar parameters in the parameter ranges they are calibrated for. However the direct relation to the mean density and surface gravity of the $\Delta\nu$ and $\nu_{\rm max}$ scaling relations are lost in this approach (see Section 2). 

Depending on the star(s) and observations of these star(s) at hand and the purpose of the stellar parameters derived using the scaling relations, the exact relation or reference function should be chosen. Certainly, one also has to be aware that both $\Delta\nu$ and $\nu_{\rm max}$ can be measured in different ways, which results in different values \citep[see e.g.,][and references therein]{hekker2011,verner2011,stello2017}, and that this should be taken into consideration when choosing a specific version of reference values or scaling relations. 

The fact that so much effort has gone into calibrating the scaling relations is testimony to the power of the $\Delta\nu$ and $\nu_{\rm max}$ scaling relations as both a simple and precise method to determine stellar parameters. With the many stars with solar-like oscillations now detected with CoROT, \textit{Kepler}, K2 and TESS, and Plato in the future, the scaling relations will provide stellar parameters for thousands of stars used in both Galactic archaeology as well as exoplanet studies, which makes the efforts discussed above worthwhile and necessary.

\section{Additional Requirements}

For additional requirements for specific article types and further information please refer to \href{http://www.frontiersin.org/about/AuthorGuidelines#AdditionalRequirements}{Author Guidelines}.

\section*{Conflict of Interest Statement}

The authors declare that the research was conducted in the absence of any commercial or financial relationships that could be construed as a potential conflict of interest.

\section*{Author Contributions}
SH has collected all the literature and wrote the manuscript. 

\section*{Funding}
SH has received funding from the European Research Council under the European Community's Seventh Framework Programme (FP7\/2007-2013) \/ ERC grant agreement no 338251 (StellarAges). 

\section*{Acknowledgments}
I would like to thank Tim Bedding and Nathalie Theme\ss l for useful comments on earlier versions of this manuscript.

\section*{Supplemental Data}
 \href{http://home.frontiersin.org/about/author-guidelines#SupplementaryMaterial}{Supplementary Material} should be uploaded separately on submission, if there are Supplementary Figures, please include the caption in the same file as the figure. LaTeX Supplementary Material templates can be found in the Frontiers LaTeX folder.

\section*{Data Availability Statement}
No data was used for this paper.

\bibliographystyle{frontiersinSCNS_ENG_HUMS} 
\bibliography{scalingreview}

\end{document}